\documentclass[11pt]{article}
\usepackage{moriond,epsfig}

\def\Journal#1#2#3#4{{#1} {\bf #2}, #3 (#4)}

\def\NPB{{\em Nucl. Phys.} B}
\def\PLB{{\em Phys. Lett.}  B}
\def\PRL{\em Phys. Rev. Lett.}
\def\PRD{{\em Phys. Rev.} D}

\def\be{\begin{equation}}
\def\ee{\end{equation}}
\def\bea{\begin{eqnarray}}
\def\eea{\end{eqnarray}}

\begin{document}
\vspace*{4cm}
\title{SELECTED RESULTS IN HADRONIC FINAL STATE PHYSICS AT HERA}

\author{ STATHES PAGANIS \\ 
(on behalf of the H1 and ZEUS Collaborations)}

\address{Columbia University, Nevis Laboratories,\\
Irvington NY, USA}

\maketitle\abstracts{
This presentation reports recent results from the hadronic 
final state in DIS at HERA. Forward jet and $\pi^0$ 
production have been measured by the H1 experiment. The forward jet 
production cross section shows significant deviation from the 
predictions based on DGLAP evolution. The forward $\pi^0$ 
data discriminate between different QCD models and are best 
described by models which take into account the partonic substructure
of virtual photons.
Inclusive $K_s^0K_s^0$ production in
$ep$ has been studied with the ZEUS detector 
using an integrated luminosity of 120 pb$^{-1}$. 
Two states are observed at masses of 
$1537$ MeV and $1726$ MeV as well as an enhancement around 
$1300$ MeV. The state at $1537$ MeV is consistent with the well 
established $f_2^\prime(1525)$. The state at $1726$ MeV may be the
glueball candidate $f_0(1710)$.
}

\section{Forward Jet and $\pi^0$ Production at HERA}
The HERA collider has extended the available kinematic range for 
Deep-Inelastic Scattering (DIS) to regions of large values of the 
four momentum trasfer $Q^2$ ($\le~10^5~GeV{^2}$) and small Bjorken-$x_{Bj}$
($x_{Bj}~\simeq~10^{-5}$). 
Studies at low $x_{Bj}$ could reveal novel features of parton 
dynamics. At small $x_{Bj}$ it is very probable that the quark 
struck by the virtual photon originates from a QCD cascade 
initiated by a parton in the proton. In different regions of 
$Q^2$ and $x_{Bj}$ different approximations to QCD are expected to describe 
the parton evolution: the most discussed being DGLAP\cite{dglap},
BFKL\cite{bfkl} and CCFM\cite{ccfm}. At high $Q^2$ and high $x_{Bj}$ the 
initial state radiation is described by the DGLAP evolution 
equations which resum the leading $\alpha_s ln(Q^2/Q^2_0)$ terms. In
this scheme a space-like chain of subsequent gluon emissions is 
characterized by a strong ordering of transverse momenta $k_T$.
However at small $x_{Bj}$ the contribution of large $ln(1/x)$ terms
may become important. Resummation of these terms leads to the BFKL
evolution equation. No ordering on transverse momenta $k_T$ of 
emitted gluons is imposed here. The CCFM evolution equation based 
on angular ordering and colour coherence interpolates between 
the BFKL and DGLAP approaches.

Differences between the different dynamical approaches to the parton
cascade are expected to  be most prominent in the 
``forward'' phase-space region
towards the proton remnant direction, i.e. away from the scattered
quark. In this talk studies of the parton evolution at small $x_{Bj}$ using
jet and $\pi^0$ production in the forward angular region in the laboratory 
frame, are presented. 
The study of DIS events with a single forward particle is
complementary to forward jet production: a forward parton 
is tagged by a single energetic fragmentation product. 

\begin{figure}
\vskip 1.5cm
\begin{center}
\epsfig{figure=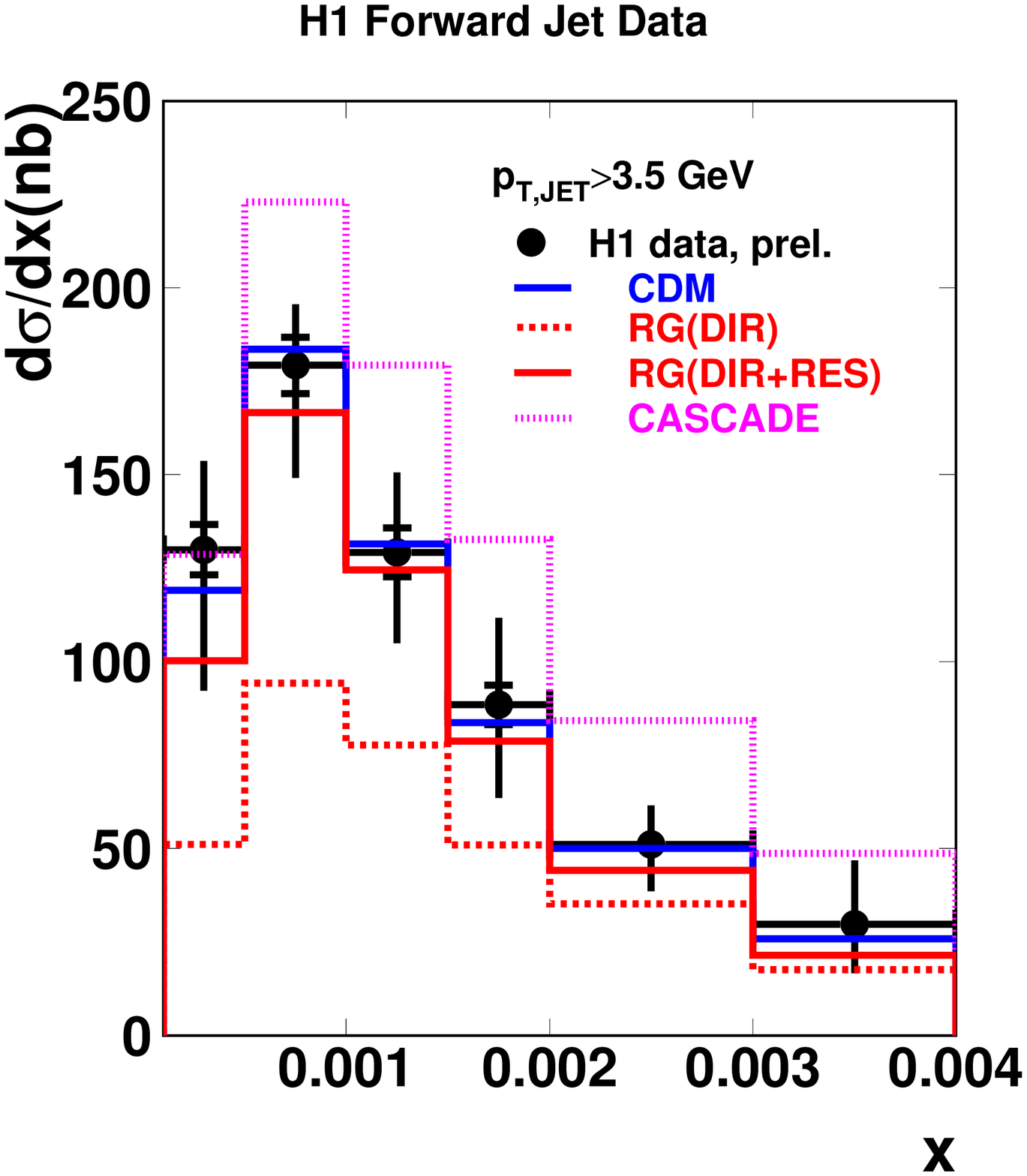,height=3.0in}
\epsfig{figure=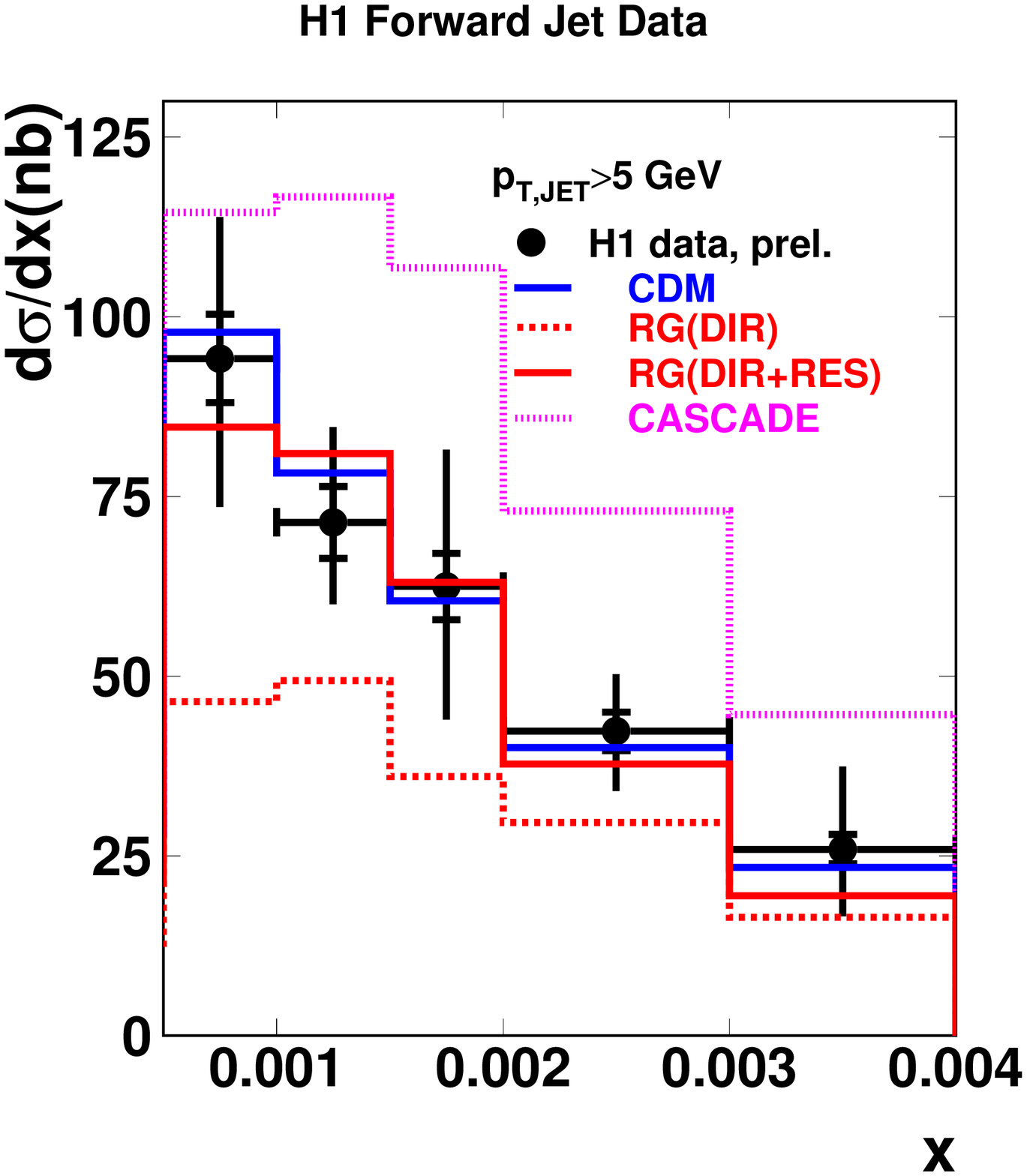,height=3.0in}
\end{center}
\caption{The cross section for forward jet production at the 
hadron level, as a function of $x_{Bj}$ 
for $p_{t~jet}>~3.5$ GeV (left) and
$p_{t~jet}>~5$ GeV (right). Also shown are the predictions from 
the CDM (ARIADNE), RAPGAP (RG) and CASCADE Monte Carlos. 
\label{fig:jets}}
\end{figure}

\begin{figure}
\vskip 1.5cm
\begin{center}
\epsfig{figure=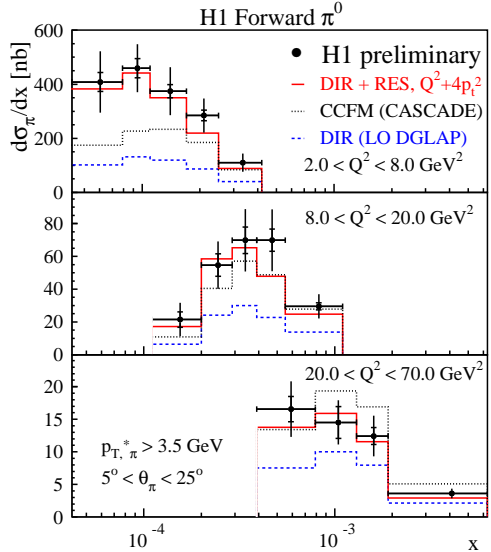,height=3.0in}
\end{center}
\caption{Inclusive $\pi^0$ production cross section as a function 
of $x_{Bj}$ for $p_{T,\pi}^*~>~3.5$ GeV in three regions of $Q^2$.
The QCD models RAPGAP based on LO DGLAP parton showers with (DIR+RES)
and without (DIR) resolved photon processes and CASCADE as an 
implementation of the CCFM equation, are compared to the data.
\label{fig:pion}}
\end{figure}

In figures~\ref{fig:jets} and~\ref{fig:pion} the cross section for 
forward jet and $\pi^0$ production as a function of $x_{Bj}$ is shown.  
The data are compared to MC model predictions. RAPGAP (RG) \cite{RAPGAP} uses
LO matrix elements supplemented with initial and final state DGLAP 
parton showers (DIR-model) and resolved virtual photon processes 
(RES-model). DJANGO \cite{DJANGO} is used together with the
Color-Dipole-Model as implemented in ARIADNE \cite{Ariadne} for 
higher order QCD radiation, labeled as CDM. In ARIADNE the parton 
emissions perform a random walk in $p_T$ leading to a situation
similar to the one expected in BFKL. CCFM evolution is implemented
in the CASCADE Monte Carlo.

In summary, for the forward jet production case
the data are up to a factor of two larger than the predicted 
cross section based on $O(\alpha_s)$ matrix elements and parton 
showers in the collinear factorization ansatz 
(DGLAP). Using a MC model incorporating resolved virtual photon processes in 
addition to the usual direct photon processes, the data are 
reasonably well described. Also the Color Dipole Model, which 
simulates higher order QCD radiation without strong ordering 
in the $p_T$ of the emitted partons, describes the measurements well. 
The CCFM approach, which is based on angular 
ordering coming from color coherence, predicts too high a rate of
forward jet events. 

Measurements of the forward $\pi^0$ 
cross-sections 
can discriminate 
between different QCD models and are best described by 
an approach in which the partonic substructure of virtual 
photons is taken into account, though with a scale of 
$Q^2+4p_T^2$ which is unusually large.

\section{Observation of $K_s^0K_s^0$ resonances in deep inelastic 
scattering at HERA}

The  $K_s^0K_s^0$ system is expected to couple to the scalar ($J^{PC}=0^{++}$) 
and tensor ($J^{PC}=2^{++}$) glueballs.
This has motivated intense 
experimental and theoretical study during the past few years \cite{review}.
Lattice QCD calculations \cite{lattice}
predict the existence of a scalar 
glueball with a mass of $1730\pm 100$ MeV and 
a tensor glueball at $2400\pm 120$ MeV.
The scalar glueball can mix with $q\overline{q}$ states with $I=0$ 
from the 
scalar meson nonet, leading to three $J^{PC}=0^{++}$ states whereas
only two fit in the nonet.
Experimentally four states with $I=0$ and $J^{PC}=0^{++}$ have been 
established: 
$f_0(980)$, $f_0(1370)$, $f_0(1500)$ and $f_0(1710)$ \cite{pdg02}.

The state most frequently considered to be a glueball candidate 
is $f_0(1710)$ \cite{pdg02} , but its gluon content has not 
yet been established. 
This state was first observed in radiative $J/\psi$ 
decays \cite{BES} and
its $J=0$ angular momentum was established by the WA102 experiment 
using a partial-wave analysis in the $K^+K^-$ and $K_s^0K_s^0$ 
final states\cite{WAf1710}.
Observation of $f_0(1710)$ in $\gamma\gamma$ 
collisions may indicate a large quark content.
A recent publication from L3 \cite{L3}  reports the observation of 
two states above $1500$ MeV in $\gamma\gamma\rightarrow K_s^0K_s^0$, 
the well established $f_2^\prime(1525)$ and a broad resonance at
$1760$ MeV. It is not clear if this state is the $f_0(1710)$.

The $ep$ collisions at HERA provide an opportunity to study resonance
production in a new environment. 
In this talk, the first observation of resonances in the $K_s^0K_s^0$ final 
state in inclusive $ep$ DIS is reported.

\begin{figure}
\vskip 1.5cm
\begin{center}
\epsfig{figure=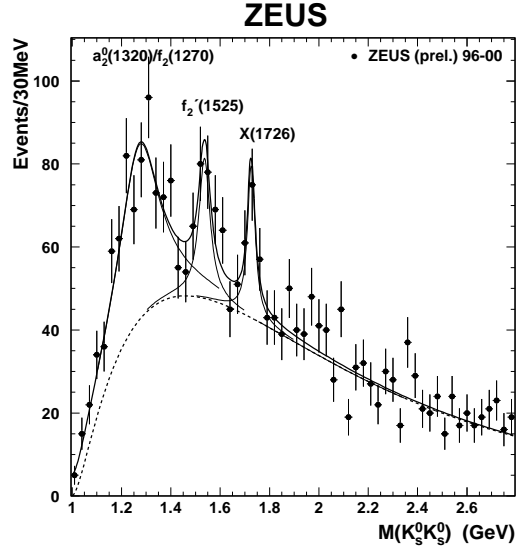,height=3.0in}
\end{center}
\caption{$K_s^0K_s^0$ invariant-mass spectrum fitted with three 
Breit-Wigners and a background function. 
The $cos\theta_{K_s^0K_s^0}$ 
cut is used to separate the $KK$ attraction
at threshold $f_0(980)$/$a_0(980)$ from the rest of the spectrum.
\label{fig:KKmass}}
\end{figure}

Oppositely charged track pairs reconstructed by the ZEUS central
tracker (CTD)
and assigned to a secondary vertex were selected and combined 
to form $K_s^0$ candidates. 
Both tracks were assigned the mass of a charged pion and the 
invariant-mass $M(\pi^+\pi^-)$ was calculated. 
Events with at least two 
$K_s^0$ candidates were selected and the 
$K_s^0K_s^0$ invariant-mass was calculated.
Figure \ref{fig:KKmass} shows the $K_s^0K_s^0$ invariant-mass spectrum 
in the range $0.995~<M(K_s^0K_s^0)~<2.795~$GeV for 
data (filled circles with error bar)
after applying a $cos\theta_{K_s^0K_s^0}<0.92$ cut.
This cut removes the effect of $f_0(980)$/$a_0(980)$ at the threshold 
which decays to collinear $K_s^0$ pairs affecting 
the measurement in the 1300 MeV mass region.
Two clear peaks are 
seen, one around 1500 MeV and the other around 1700 MeV.
This is the first observation in $ep$ DIS of a state 
near $1537$ MeV consistent with $f^\prime_2(1525)$ and
another near $1726$ MeV close to $f_0(1710)$. There
is also an enhancement near $1300$ MeV which may arise from the 
production of $f_2(1270)$/$a_2^0(1320)$.


\section*{References}
\begin{thebibliography}{99}
\bibitem{dglap}
V.N. Gribov, L.N. Lipatov, {\it Sov.J.Nucl.Phys.} {\bf 15}, 438 and 675
(1972);
Yu.L. Dokshitzer, {\it Sov. Phys. JETP} {\bf 46}, 641 (1977);
G. Altarelli, G. Parisi, {\it Nucl. Phys.} {\bf 126}, 297 (1978). 
\bibitem{bfkl}
E.A. Kuraev, L.N. Lipatov, V.S. Fadin, {\it Sov. Phys. JETP} {\bf 45}, 
199 (1972);
Y.Y. Balitsky, L.N. Lipatov, {\it Sov.J.Nucl.Phys.} {\bf 28}, 822 (1978).
\bibitem{ccfm}
M. Ciafaloni, \NPB {\bf~296}, 49 (1988);
S. Catani, F. Fiorani, G. Marchesini, \Journal{\PLB}{234}{339}{1990}, 
\NPB {\bf~336}, 18 (1990);
G. Marchesini, \NPB {\bf~445}, 49 (1995).
\bibitem{RAPGAP}
H. Jung, {\sf http://www.quark.lu.se/$\sim$hannes/rapgap/}.
\bibitem{DJANGO}
G.A. Schuler, H. Spielberger, {\it Proc. Physics at HERA}, 
vol.3, 1419-1432, Hamburg 1991.
\bibitem{Ariadne}
L. Lonnblad, {\it Comp. Phys. Comm.} {\bf 71}, 15 (1992).
\bibitem{review}
S. Godfrey, J. Napolitano, {\it Review of Mod. Phys.} {\bf 71}, 1411 (1999).
\bibitem{lattice}
C.J. Morningstar, M. Peardon, \Journal{\PRD}{60}{034509}{1999};
C. Michael, M. Teper, \Journal{\NPB}{314}{347}{1989}.
\bibitem{pdg02}
K. Hagiwara et al, \Journal{\PRD}{66}{1}{2002}.
\bibitem{BES}
J.Z. Bai et al,  \Journal{\PRL}{77}{3959}{1996}.
\bibitem{WAf1710}
D. Barberis et al,  \Journal{\PLB}{453}{305}{1999}.
\bibitem{L3}
M. Acciari et al,  \Journal{\PLB}{501}{173}{2001};

\end{thebibliography}
\end{document}